# Intriguing uniform elongation and accelerated radial shrinkage in amorphous SiO$_x$ nanowire as purely induced by uniform electron beam irradiation

Jiangbin Su[1,2,†], Xianfang Zhu[1,3*†]

1. *China-Australia Joint Laboratory for Functional Nanomaterials & Physics Department, Xiamen University, Xiamen 361005, People's Republic of China*

2. *Experiment Center of Electronic Science and Technology, School of Mathematics and Physics, Changzhou University, Changzhou 213164, People's Republic of China*

3. *Institute of Biomimetics and Soft Matter, Xiamen University, Xiamen 361005, People's Republic of China*

\* Corresponding author. E-mail: zhux@xmu.edu.cn

† Both authors contributed equally to this work and should be considered as co-first authors.

**Abstract**: An intriguing athermal uniform elongation and accelerated radial shrinkage in amorphous SiO$_x$ nanowire (a-SiO$_x$ NW) as purely induced by uniform electron beam (e-beam) irradiation were in-situ investigated in transmission electron microscope. It was observed that at room temperature the straight and uniform a-SiO$_x$ NW demonstrated a uniform plastic elongation (without any external tensile pulling) and a corresponding uniform radial shrinkage. According to a new model taking into account the NW nanocurvature over its surface as well as the beam-induced athermal activation, a kinetic relationship between the shrinking radius and the irradiation time was established. The fitting results demonstrated that a curvature-dependent surface energy at the nanoscale (so-called nanocurvature effect) much higher than that predicted from the existing theories is exerted on the elongation and shrinkage of the NW. At the same time, the so-induced plastic flow of massive atom and surface diffusion of atom presented as well a further direct experimental evidence for our predicted soft mode and instability of atomic vibration as induced under energetic beam (including e-beam) irradiation in amorphous materials The study has important implications for the nanoprocessing or nanostability of future NW-based structures or devices. More importantly, it further demonstrates that the nanocurvature effect and the beam-induced atomic vibration soft mode and instability effect, which have been normally neglected or inadequately taken into account in the current literature, are universal concepts and applicable to explanation of energetic beam-induced nanoinstability or nanoprocessing of low dimensional nanostructure in general.

## 1. Introduction

Electron beam (e-beam) irradiation-induced structure changes and processing in nanowires (NWs) have been studied extensively [1-10] due to their intensive research interest and numerous promising applications in nanotechnology and nanodevices. Among these, study on the elongation of NWs [7-10] in an *in-situ* transmission electron microscope (TEM) is especially popular and interested. This is because such a study makes great sense not only to the fundamental science of plastic transition but also to the widening of practical applications of NWs, which may be limited by the difficulties in controllable fabrication of NWs with desired shapes such as length and diameter. Up till now, Zheng *et al* [7], Han *et al* [8] and Zheng *et al* [9] have reported the low-intensity (non-focused) e-beam-assisted tensile pulling-induced elongation of crystalline $Li_2O$ and silicon NWs and amorphous silica NW respectively, in which both the elongation mechanisms and stretching installations were complicated.

Furthermore, Zhu *et al* [10] have irradiated amorphous $SiO_x$ (a-$SiO_x$) NW by a focused e-beam, where a local S-type prolonged and necked NW was realized. Even so, the elongation and necking kinetics were complicated due to the non-uniform distributions of both positive and negative nanocurvatures as well as of the beam current density over the necked segment with the possible refilling of atoms into the necked segment [10]. Therefore, we expect to track the kinetics of uniform axial elongation and radial shrinkage to correlate accurately our modeling of nanocurvature effect [11-12] *via* a uniform irradiation with a beam spot size large enough over a whole NW. Such a NW must be straight and uniform in diameter all the way in the axis direction with a surface smooth enough to avoid the effect of any non-uniformity of the nanocurvature on the sidewall of wire. Moreover, such a wire must be thin enough from the very beginning to ensure a shrinkage to a few nanometers in diameter where within a feasible irradiation duration, during which the beam current can be kept high enough without any appreciable fluctuation. Actually, it has been proven rather difficult to obtain such a beam or such a NW [10]. To our best knowledge, up till now, there has no report been found on the uniform elongation and radial shrinkage of NWs as purely induced by a uniform irradiation of e-beam. Hence, the nanocurvature effect [11-12] and the beam-induced atomic vibration soft mode and instability (or beam-induced athermal activation) effect [11,13] underlying the as-induced uniform elongation and radial shrinkage of NWs remain unclear.

Based on the above considerations, in this paper we particularly study the uniform elongation and radial shrinkage of a-$SiO_x$ NWs as purely induced by a uniform irradiation of non-focused e-beam in an *in-situ* TEM. It was observed that the straight and uniform a-$SiO_x$ NW presented a tensile pulling-free uniform plastic elongation and an accelerated uniform radial shrinkage at the nanoscale intriguingly. According to our new model taking into account the NW curvature at the nanoscale

(nanocurvature), the kinetic relationship between the shrinking radius and the irradiation time was estimated. The fitting results showed a much higher, curvature-dependent surface energy at the nanoscale than that predicted from the existing theories on the shrinkage (that is, a nanocurvature effect on the shrinkage). Furthermore, the arresting plastic flow of massive atoms and surface diffusion of atoms at room temperature indicate a direct experimental evidence for our predicted athermal soft mode and instability of atomic vibration as induced under non-focused, uniform e-beam irradiation in condensed matter.

## 2. Experimental

The a-SiO$_x$ NWs with the desired diameter were fabricated by our improved chemical vapor deposition technique where x is determined to be 2.3 [14]. Such fabricated NWs can be straight and uniform in diameter all the way in the axis direction with a surface smooth enough to avoid the effect of any non-uniformity of the nanocurvature on the sidewall of wire. The as-fabricated NWs were well-dispersed in ethanol and then deposited onto the holey carbon film of microscopy grid for TEM studies. The as-prepared TEM specimens were irradiated at room temperature and the structure changes and evolution of SiO$_x$ NWs was examined using our developed *in-situ* TEM observation technique *via* a field-emission Tecnai F-30 TEM operating at a high tension of 300 kV. The irradiation was always targeted on individual straight NW segments with uniform diameter and clean and smooth surface, which were protruding into the open space of the holes in the carbon film of microscopy grid. In each irradiation, the beam current density at the specimen could be high enough without any appreciable fluctuation and was kept at about 5.66 A/cm$^2$ (flux: 3.54×10$^5$ nm$^{-2}$s$^{-1}$), which was uniform over an area (about 600 nm in diameter) much larger than the zone or NW segment observed (about 200 nm in length). During the observation or taking a picture, the beam was spread to an around 100 times weaker intensity so that the corresponding irradiation effect can be minimized to a negligible degree and at the same time the image contrast can also be improved. Also note that during the electron irradiation, the beam was determined to heat the specimen by no more than a few degrees [10,15-16] due to its extremely large ratio of surface to volume and the dominant irradiation effect should be athermal. Therefore, it could be considered that the irradiated NW essentially remained at room temperature throughout the irradiation duration. In the experiments, the wire diameter was taken as an average value of several wire diameters which are measured at different representative positions across the wire from each micrograph; while the length and volume of the wire segment were measured and calculated between two red mark dots in Fig. 1 along the wire axis. The two red dots were carefully marked at the locations of two feature points and by further checking the relative positions of their surrounding feature points on the wire surface (see the arrows in Fig. 1).

## 3. Results and discussions

The sequential TEM micrographs in Fig. 1 show the typical structure changes and evolution of an individual a-SiO$_x$ NW segment during the uniform e-beam irradiation at room temperature. Prior to the irradiation, as shown in Fig. 1(a), the a-SiO$_x$ NW segment within the two red mark dots has a well-defined straight, smooth, and clean cylinder shape with an initial length of 148 nm and its two ends lying on the supporting carbon film. With the irradiation turned on, as demonstrated in Figs. 1(a-k) and 2, sequential consistent extension in the segment length after each dose of the beam was occurring intriguingly. Furthermore, the segment almost kept its straight and smooth cylinder shape, up till a length about 192 nm for the marked wire segment was finally attained without breakage, as shown in Fig. 1(k). Such uniform axial elongation from 148 nm to 192 nm within 3540 s irradiation demonstrates a total extension in the segment length of 44 nm and an average extension rate of

1.24×10$^{-2}$ nm/s. As compared with the initial NW segment, there is an increment up to 30% in the length without the external assistance of tensile pulling [7-9]. It is expected that such uniform elongation of a NW segment should be always accompanied with a uniform radial shrinkage. As experimentally demonstrated in Figs. 1(a-k), a notable uniform shrinkage in the wire diameter was indeed observed from 19 nm to 8 nm, which finally led the wire to breakage when irradiated to 3600 s (see Fig. 1(l)). Fig. 3 further shows the plot of radius of the NW segment in Fig. 1 against the irradiation time. It was observed that the shrinkage started to become faster and faster intriguingly when the radius reduced down to about 7 nm. In the existing literature, such an accelerated uniform radial shrinkage at the nanoscale under uniform e-beam irradiation has not been observed in NWs, which should be greatly different from the cases of single-walled carbon nanotubes (SWCNTs) [17] or nanocavities in Si [11,18] due to the unique amorphous NW structure. A similar uniform irradiation on other NW segments by changing the beam current densities down to 1 A/cm$^2$ was repeated several times. It was observed that the features of the structure changes and evolution were essentially the same as shown in Fig. 1 although the elongation and radial shrinkage were both slower and less-proceeded at a lower current density. We also note that there is still a little coarseness on the NW segment surface after a long period of irradiation, as shown in Fig. 1. This is probably due to the minor non-uniformity or fluctuation either in local beam current density or in local wire structure or in both during the irradiation, which seems to be inevitable at a prolonged, later-on irradiation stage.

According to conventional views, the main mechanisms of interaction between energetic e-beam and solid materials are often simply attributed to knock-on mechanism [16] or even e-beam heating effect [1,19-20]. However, the knock-on mechanism and its related simulations neither are fully consistent with, nor can offer a full explanation for, the experimentally observed wire structure changes and evolution, especially the above tensile pulling-free uniform elongation and the accelerated radial shrinkage at the nanoscale during the irradiation. This is because the existing theories such as the knock-on mechanism and their related simulations were at the first place built on consideration of the nature of equilibrium, symmetry, periodicity, and linearity of bulk crystalline structure or its approximation, whereas the energetic beam (including electron, ion and photon beams)-induced nanophenomena are intrinsically of non-equilibrium, amorphous, and non-linear nature. Moreover, the atomic movements at the scale of atomic bond length are difficult to be observed with the current TEM techniques, and thus a direct comparison of the simulated atomic movements of the wires with the experimental results is impossible. On the other hand, as mentioned in the experimental part, due to the extremely high ratio of surface to volume of NW at the nanoscale, the e-beam irradiation is determined to heat the specimen by no more than a few degrees [10,15-16] and the dominant irradiation effect should be athermal. In the experiment, we also observed that even after the above elongation and shrinkage processed for a time, they would stop immediately once the irradiation was suspended. This further demonstrates that the processes are predominately driven by a fast, irradiation-induced athermal activation rather than a slow, beam heating-induced thermal activation. In fact, our previous work on the energetic beam irradiation-induced structure changes and processing of low dimensional nanostructures (LDNs), such as nanocavities in Si [11,18,21-23], carbon nanotubes [17,24-25] and amorphous $SiO_x$ nanowires [6,10] has proven that our proposed novel nanocurvature effect [11-12] and energetic beam-induced atomic vibration soft mode and instability effect [11,13] are universal concepts and applicable in the prediction or explanation of all the above energetic beam irradiation-induced nanophenomena. In the following, we attempt to reveal the above two effects on the uniform elongation and the accelerated uniform radial shrinkage at the nanoscale of a-SiO$_x$ NW under uniform e-beam irradiation.

For the nanocurvature effect on a NW, we can suppose that, similar to the particle case [11-12], when the radius *r* of a NW approaches its atomic bond length *d*, a positive nanocurvature on the highly curved wire surface will become appreciable. Such a positive nanocurvature would cause an additional tensile stress on the electron cloud structure of surface atoms which would lead to a dramatic increase in surface energy of the NW as schematically shown in Fig.4(a). This dramatically increased surface energy would give rise to a strong tendency of self-shrinkage on the nanocurved wire surface and thus provide a thermodynamic driving force for the wire to contract in the radius direction and massive atom plastic flow as self-extruded towards two ends of the wire.

Although the positive nanocurvature on a NW can cause a strong tendency of self-shrinkage on the wire surface, a further assistance from the external excitation such as energetic beam irradiation is still needed to soften the wire and kinetically activate the shrinkage or contraction and the plastic flow. This can be verified by the observation that the elongation and radial shrinkage of the NWs would stop immediately once the irradiation was suspended even after the structure changes started. In the present case of energetic e-beam irradiation in TEM, we can assume that beam energy deposition rate [6] of the incident energetic beam can be so fast that there is no enough time for the deposited energy to transfer to thermal vibration energy of atoms within a single period of the vibration. In this way, the mode of atom thermal vibration would be softened or the vibration of atoms would lose stability [11,13]. The induced soft mode or instability of the atomic vibration can suppress the energy barrier for the atoms to migrate or escape and finally make the structure changes kinetically possible.

The above processes can be schematically illustrated in Fig. 4(b). When the NW segment is subjected to a uniform irradiation of e-beam, the mode of the atom thermal vibration is softened or the vibration of atoms lose stability within the whole irradiated NW segment. As a result, the energy barriers of atoms in the NW segment were suppressed, and the atoms became unstable and easy to diffuse or even flow towards two ends of the wire or evaporate into the vacuum under effect of the self-extruding of the wire nanocurved-surface. Meanwhile, since the nanocurvature distributed over the wire surface is also uniform along the wire axis, the as-induced athermal diffusion and evaporation of surface atoms and plastic flow of massive atoms would proceed uniformly along the wire axis. In this way, as demonstrated in Fig. 1 and further illustrated in Fig. 4(b), a uniform axial elongation and a uniform radial shrinkage were occurring in the a-SiO$_x$ NW during the irradiation of a uniform e-beam. Thus, the novel plastic flow of massive atoms and surface diffusion of atoms offer a direct experimental evidence for our predicted athermal soft mode and instability of atomic vibration as induced under uniform e-beam irradiation in condensed matter. In particular, with the increase of irradiation time, the NW became thinner and demonstrated an intriguing accelerated radial shrinkage at the nanoscale as shown in Fig. 3. This can be attributed to the nanocurvature effect of the NW when the NW become thinner with a radius down to the nanoscale.

To reveal the kinetics of the accelerated radial shrinkage during the uniform elongation, we can consider an irradiated NW segment of any fixed length of $L$, as illustrated in Fig. 4 and correlated the thermodynamic driving force for the radius shrinkage with the surface energy (capillary force) of the wire through Laplace law

$$F = (2\pi r L)\sigma/r, \qquad (1)$$

where $F$ represents the driving force, $r$ and $\sigma$ are the radius and the surface energy of the wire respectively, and $2\pi rL$ is the surface area of the wire segment. From classical thermodynamics, the surface energy of a wire is independent of the curvature $1/r$ or radius $r$ (i.e., $\sigma = \sigma_o$). However, when a wire radius become comparable to the atomic bond length at the nanoscale, an additional nanocurvature effect on the surface energy emerges, which depends on the radius (i.e., $\sigma$ increases dramatically with the decreasing of $r$) and can be assumed to be proportional to $\alpha/r^b$, for example. Thus, we can have

$$\sigma = \sigma_0 + \alpha/r^b, \qquad (2)$$

where $\alpha$ is introduced as surface energy factor and $b$ is considered as a constant ($b \geq 0$). For the a-SiO$_x$ NW with a nanoscale radius, the additional surface energy ($\alpha/r^b$) as caused by additional nanocurvature effect becomes very large (i.e., $\alpha/r^b >> \sigma_o$), thus the surface energy

$$\sigma \approx \alpha/r^b. \qquad (3)$$

From kinetic consideration, we can further assume that rate of the mass transportation (decreasing number of atoms within the irradiated NW segment of the fixed length of $L$) during the elongation and the shrinkage is proportional to the driving force as shown in Eq. (1) [18]. Note that the decreased number of the atom includes both the number of atoms diffused (or flowed) out of and that evaporated out of the irradiated NW segment of the fixed length of $L$. In this way, we can get

$$-\mathrm{d}(\pi r^2 L \rho)/\mathrm{d}t \propto F, \qquad (4)$$

or

$$-\mathrm{d}(\pi r^2 L \rho)/\mathrm{d}t = K F. \qquad (5)$$

In Eq. (4) or (5), $\rho$ is the volume density of atoms which is taken as a constant value here, and $K$ is the thermodynamic reaction constant which is defined as $K = K_0 \exp(-\Delta G^*/RT)$, where $K_0$ is a constant, $R$ is the gas constant, $T$ is the temperature, and $\Delta G^*$ is the activation energy for the structure changes [17]. Combining Eqs. (1), (3), and (5), we then have

$$r^{b+1}\, dr = -K'\, dt, \qquad (6)$$

where $K' = K\alpha/\rho$. Before irradiation, the wire radius is $R_0$, whilst after irradiating the wire to time $t$, the radius becomes $r$

$$r_{t=0} = R_0, \quad r_{t=t} = r(t). \qquad (7)$$

By integrating the Eq. (6) and employing the conditions given in Eq. (7), $r$ takes the form as given below

$$r = [R_0^\eta - \mu\, t]^{\frac{1}{\eta}}, \qquad (8)$$

where $\eta = b+2$ and $\mu = K'(b+2)$. Thus, the Eq. (8) establishes a kinetic relationship between the shrinking wire radius and the e-beam irradiation time, which could reveal the physical insights into the structure changes and evolution of the NW under the e-beam irradiation.

Based on the above considerations, we made a nonlinear fitting of the observed radius evolution of the NW as shown in Fig. 3 to the Eq. (8). The dashed line in Fig. 3 shows a well-fitted result of the experimental data which gives a fitted value of the constant $b$ at 3.5. This indicates that the additional curvature effect at the nanoscale on the surface energy of the NW is proportional to $1/r^{3.5}$ instead of $1/r^2$ as predicted from the existing theories [17]. That is, the fitting results demonstrate a much faster increase of the surface energy with the curvature at the nanoscale than that predicted from the existing theories. There may be two reasons for such a discrepancy: first, the existing theories have only predicted the curvature energy at a static condition without being subject to any e-beam irradiation; second, the related, existing calculations must rely on some approximations based on equilibrium, symmetric, periodic, and linear nature of a crystalline structure, which may greatly underestimate real curvature effect of LDNs at the nanoscale. Therefore, when the radius $r$ of wire reduces down to a value comparable with the atomic bond length, the surface energy increases dramatically in the form of $\alpha/r^{3.5}$, which could cause a strong tendency of self-compression on the nanocurved wire surface thermodynamically. Further, under the external activation of the uniform e-beam irradiation, the dramatically increased surface energy can athermally drive the atom diffusion or even plastic flow of massive atoms near surface towards the two ends of wire as well as the athermal surface atom evaporation into the vacuum. As a result, the a-SiO$_x$ NW shows a tensile pulling-free plastic elongation and an accelerated radial shrinkage at the nanoscale under the uniform e-beam irradiation, as demonstrated in Figs. 1-3.

In our previous work, we have obtained the constant $b$ at 6 for SWCNT [17] which is larger than that in the present case of a-SiO$_x$ NW with a fitted value of 3.5. This can be well accounted for from the great differences of structure configurations or nanocurvature effects between the SWCNT and the a-SiO$_x$ NW. Firstly, for the SWCNT, it can be regarded as a hollow tube structure rolled up from an sp$^2$ monoatomic layer of covalent bond with an outer surface and an inner surface. As a result, both a strong tensile stress by a positive curvature on the outer surface of the tube wall and a strong compressive stress by a negative curvature on the inner surface would lead to the dramatic increase in the surface energy of SWCNT. As shown in the Eq. (1) in ref. [17], there are two times the surface energy (i.e. 2$\sigma$) in the expression of the driving force. While for the a-SiO$_x$ NW, it is an amorphous solid cylinder structure of Si-O-Si bridge bond with only an outer wire surface where only the tensile stress by a positive curvature could cause the increase in the surface energy. As shown in the Eq. (1), there is only one time the surface energy in the expression of the driving force. Secondly, the number of atoms in the SWCNT and a-SiO$_x$ NW segments of a length at $L$ are respectively proportional to

($2\pi rL$) and ($\pi r^2L$), as shown in Eqs. (4-8), which would finally lead to a smaller value of the constant $b$ for a-SiO$_x$ NW (SWCNT: $b = \eta$-1 [17]; a-SiO$_x$ NW: $b = \eta$-2). Lastly, the initial radius of a-SiO$_x$ NW in this paper is much larger than that of SWCNT in ref. [17]: 9.5 nm for a-SiO$_x$ NW and 0.9 nm for SWCNT. All the above indicate a less-notable nanocurvature effect and a lower surface energy of a-SiO$_x$ NW, which supports the conclusion of a smaller constant $b$ in the a-SiO$_x$ NW relative to the SWCNT. In addition, both the smaller current density (a-SiO$_x$ NW: 5.66 A/cm$^2$; SWCNT: 100 A/cm$^2$) and the much more materials for a-SiO$_x$ NW, in which the beam energy deposition rate should be much lower according to the Eq. (1) in ref. [6], may also relax the strong tendency of shrinking in radius.

## 4. Conclusions

In summary, the structure changes and evolution of a-SiO$_x$ NW under uniform e-beam irradiation were *in-situ* investigated in TEM. It was observed that the straight and uniform a-SiO$_x$ NW intriguingly manifests itself by a tensile pulling-free uniform plastic elongation and an accelerated uniform radial shrinkage at the nanoscale. According to our new model taking into account the NW curvature at the nanoscale, the kinetic relationship between the shrinking radius and the irradiation time was established. The fitting results showed a much higher curvature-dependent surface energy at the nanoscale than that predicted from the existing theories on the shrinkage. Furthermore, the plastic flow of massive atoms and surface diffusion of atoms indicate a direct experimental evidence for our predicted athermal soft mode and instability of atomic vibration as induced under uniform e-beam irradiation in condensed matter. It is expected that such a study has important implications for the nanoprocessing or nanostability of future NW-based structures or devices. More importantly, it further demonstrates that the novel nanocurvature effect and beam-induced atomic vibration soft mode and instability effect, which have been normally neglected or inadequately taken into account in the current literature, are able to explain the beam-induced NW structure changes and evolution. Thus, the new concepts could offer a mechanism different from the classical knock-on mechanism and e-beam heating effect and thus may well reveal the fundamentals of the beam-induced nanophenomena in NWs of non-equilibrium, amorphous, and non-linear nature.

## Acknowledgements

We acknowledge the support in the TEM observation from Qiming Hong at Xiamen University. This work was supported by the NSFC project under grant no. 11574255, the Science and Technology Plan (Cooperation) Key Project from Fujian Province Science and Technology Department under grant no. 2014I0016, and the National Key Basic Science Research Program (973 Project) under grant no. 2007CB936603.

## References

[1] Xu S, Tian M, Wang J, Xu J, Redwing JM and Chan MHW. Nanometer-scale modification and welding of silicon and metallic nanowires with a high-intensity electron beam. Small, 2005, 1: 1221-1229.

[2] Remeika M and Bezryadin A. Sub-10 nanometer fabrication: molecular templating, electron-beam sculpting and crystallization of metallic nanowires. Nanotechnology, 2005, 16: 1172-1176.

[3] Pecora EF, Irrera A, Boninelli S, Romano L, Spinella C and Priolo F. Nanoscale amorphization, bending and recrystallization in silicon nanowires. Applied Physics A, 2011, 102: 13-19.

[4] Dai S, Zhao J, Xie L, Cai Y, Wang N and Zhu J. Electron-beam-induced elastic-plastic transition in Si nanowires, Nano Letters, 2012, 12: 2379-2385.

[5] Dai S, He M and Zhu J. E-beam-induced in situ structural transformation in one-dimensional nanomaterials, Science Bulletin, 2015, 60: 71-75.

[6] Su J and Zhu X. Atom diffusion and evaporation of free-ended amorphous $SiO_x$ nanowires: nanocurvature effect and beam-induced athermal activation effect, Nanoscale Research Letters, 2016, 11: 514.

[7] Zheng H, Liu Y, Mao SX, Wang J and Huang JY. Beam-assisted large elongation of in situ formed $Li_2O$ nanowires. Scientific Reports, 2012, 2: 1-4.

[8] Han X, Zheng K, Zhang Y, Zhang X, Zhang Z and Wang ZL. Low-temperature in situ large-strain plasticity of silicon nanowires. Advanced Materials, 2007, 19: 2112-2118.

[9] Zheng K, Wang C, Cheng YQ, Yue Y, Han X, Zhang Z, Shan Z, Mao SX, Ye M, Yin Y and Ma E. Electron-beam-assisted superplastic shaping of nanoscale amorphous silica. Nature Communications, 2010, 1: 24.

[10] Zhu X, Su J, Wu Y, Wang L and Wang Z. Intriguing surface-extruded plastic flow of $SiO_x$ amorphous nanowire as athermally induced by electron beam irradiation, Nanoscale, 2014, 6: 1499-1507.

[11] Zhu X and Wang Z. Nanoinstabilities as revealed by shrinkage of nanocavities in silicon during irradiation, International Journal of Nanotechnology, 2006, 3: 492-516.

[12] Zhu X. Evidence of antisymmetry relation between a nanocavity and a nanoparticle: a novel nanosize effect, Journal of Physics: Condensed Matter, 2003, 15: L253-L261.

[13] Zhu X and Wang Z. Evidence of ultrafast energy exchange-induced soft mode of a phonons and lattice instability: a nanotime effect, Chinese Physics Letters, 2005, 22: 737-740.

[14] Huang S, Wu Y, Zhu X, Li L, Wang Z, Wang L and Lu G. VLS growth of $SiO_x$ nanowires with a stepwise nonuniformity in diameter. Journal of Applied Physics, 2011, 109: 084328.

[15] Zhang J, You L, Ye H and Yu D. Fabrication of ultrafine nanostructures with single-nanometer precision in a high-resolution transmission electron microscope. Nanotechnology, 2007, 18: 155303.

[16] Banhart F. Irradiation effects in carbon nanostructures. Reports on Progress in Physics, 1999, 62: 1181-1221.

[17] Zhu X, Li L, Huang S, Wang Z, Lu G, Sun C and Wang L. Nanostructural instability of single-walled carbon nanotubes during electron beam induced shrinkage. Carbon, 2011, 49: 3120-3124.

[18] Zhu X. Shrinkage of nanocavities in silicon during electron beam irradiation. Journal of Applied Physics, 2006, 100: 034304.

[19] Ugarte D. Curling and closure of graphitic networks under electron-beam irradiation, Nature, 1992, 359: 707-709.

[20] Kiang CH, Goddard WA, Beyers R and Bethune DS. Structural modification of single-layer carbon nanotubes with an electron beam, The Journal of Physical Chemistry, 1996, 100: 3749-3752.

[21] Zhu XF, Williams JS, Llewellyn DJ and McCallum JC. Instability of nanocavities in amorphous silicon, Applied Physics Letters, 1999, 74: 2313-2315.

[22] Zhu XF, Williams JS, Conway MJ, Ridgway MC, Fortuna F, Ruault MO and Bernas H. Direct observation of irradiation-induced nanocavity shrinkage in Si, Applied Physics Letters, 2001, 79: 3416-3418.

[23] Zhu X and Wang Z. Nanocavity shrinkage and preferential amorphization during irradiation in silicon, Chinese Physics Letters, 2005, 22: 657-660.

[24] Zhu X, Li L, Su J and Wang L. Beam-induced nonuniform shrinkage of single-walled carbon nanotube and passivation effect of metal nanoparticle. Journal of Physical Chemistry C, 2015, 119: 6239-6245.

[25] Zhu X, Gong H, Yang L, Li L and Sun C. Non uniform shrinkages of double-walled carbon nanotube as induced by electron beam irradiation. Applied Physics Letters, 2014, 105: 093103.

# Figure captions

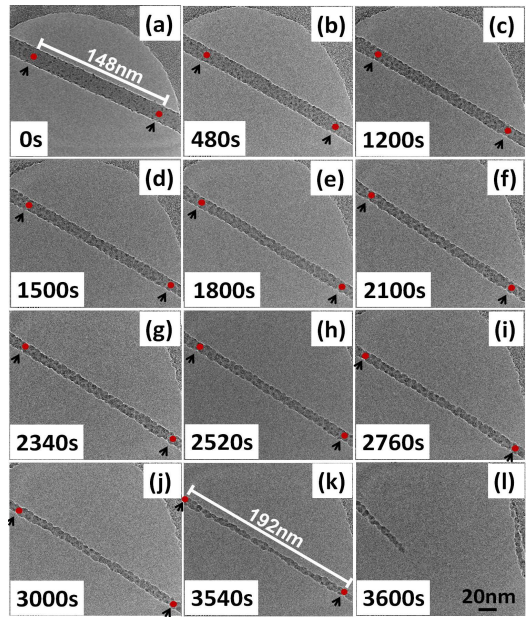

**Fig. 1** Sequential TEM images (a-l) showing the uniform structure changes and evolution with the irradiation time as induced by irradiation of a uniform e-beam (with current density of ~5.66A/cm$^2$) on an a-SiO$_x$ NW (of ~19nm in wire diameter).

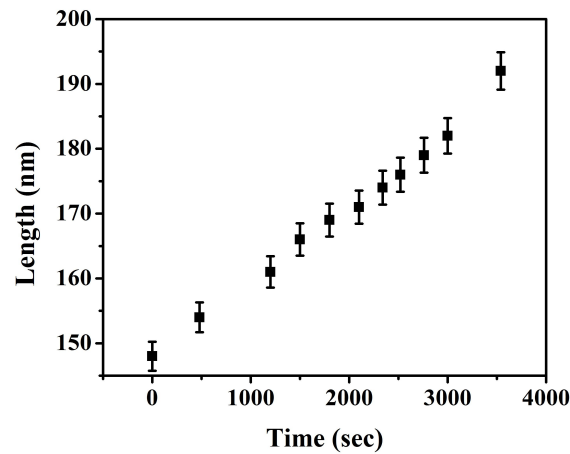

**Fig. 2** Changes of length of the NW with the irradiation time as measured from Fig. 1. The length is measured as the length of the irradiated wire segment between the two red feature dots as shown in Fig. 1 to trace the changes of wire length.

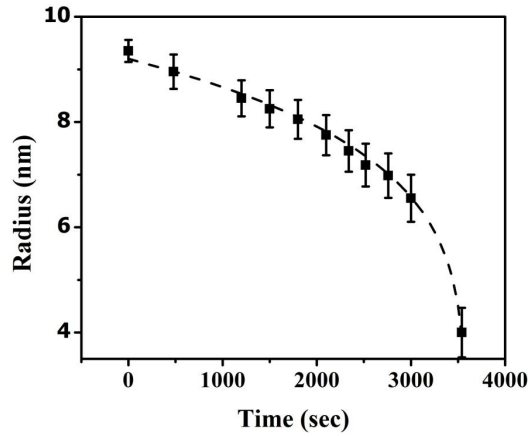

**Fig. 3** The uniform shrinkage of the SiO$_x$ NW radius ($r$) with the irradiation time ($t$) as observed in Fig. 1. The dashed line is the fitting result according to the Eq. (8) with a value of the constant $b$ at about 3.5.

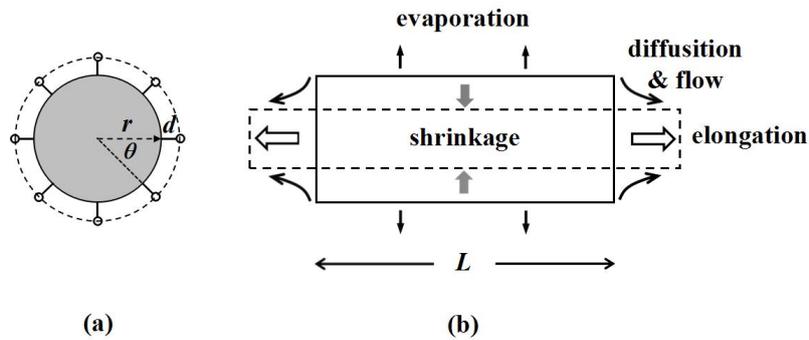

**Fig. 4** Schematic illustration showing (a) the surface nanocurvature effect of a-SiO$_x$ NW, and (b) the mechanisms of uniform axial elongation and radial shrinkage of a wire segment of any fixed length of $L$ as taken from a-SiO$_x$ NW irradiated under uniform e-beam irradiation.